\documentclass[journal]{IEEEtran}
\IEEEoverridecommandlockouts
\usepackage{cite}
\usepackage{amsmath,amssymb,amsfonts}
\usepackage{minipage-marginpar}
\usepackage{enumitem}
\usepackage{graphicx}
\usepackage{textcomp}
\usepackage{xcolor}
\usepackage{url}            
\usepackage{booktabs}       
\usepackage{amsfonts}       
\usepackage{mathtools}
\usepackage{float}
\usepackage{dblfloatfix}    
\usepackage{multirow}
\usepackage{algpseudocode}
\usepackage[ruled,vlined,linesnumbered]{algorithm2e}

\usepackage{amsthm}
\usepackage{etoolbox}

\AfterEndEnvironment{theorem}{\noindent\ignorespaces}

\AfterEndEnvironment{corollary}{\noindent\ignorespaces}

\AfterEndEnvironment{lemma}{\noindent\ignorespaces}

\AfterEndEnvironment{definition}{\noindent\ignorespaces}

\AfterEndEnvironment{example}{\noindent\ignorespaces}
\usepackage{amssymb}
\usepackage{amsmath}
\DeclareMathOperator*{\argmax}{arg\,max}
\DeclareMathOperator*{\argmin}{arg\,min} 

\usepackage{xfrac}
\usepackage{array}
\newcolumntype{x}[1]{>{\centering\arraybackslash\hspace{0pt}}p{#1}}

\def\BibTeX{{\rm B\kern-.05em{\sc i\kern-.025em b}\kern-.08em
		T\kern-.1667em\lower.7ex\hbox{E}\kern-.125emX}}
\begin{document}

%
\title{Epidemic Exposure Notification with Smartwatch:\\
A Proximity-Based Privacy-Preserving Approach}

\author{Pai~Chet~Ng,~\IEEEmembership{Student~Member,~IEEE,}
    Petros~Spachos,~\IEEEmembership{Senior~Member,~IEEE,} 
    Stefano~Gregori,~\IEEEmembership{Senior~Member,~IEEE,} and 
    Konstantinos~N.~Plataniotis,~\IEEEmembership{Fellow,~IEEE}
    \thanks{Pai Chet Ng is with the Department of Electronics and Computer Engineering, Hong Kong University of Science and Technology, Hong Kong. E-mail: pcng@ust.hk}
    \thanks{Petros Spachos and Stefano Gregori are with the School of Engineering, University of Guelph, Canada. E-mail: \{petros, sgregori\}@uoguelph.ca}%
    \thanks{Konstantinos N. Plataniotis is with the Department of Electrical and Computer Engineering, University of Toronto, Canada. E-mail: kostas@ece.utoronto.ca}
}



\maketitle
\begin{abstract}
Businesses planning for the post-pandemic world are looking for innovative ways to protect the health and welfare of their employees and customers. Wireless technologies can play a key role in assisting contact tracing to quickly halt a local infection outbreak and prevent further spread. In this work, we present a wearable proximity and exposure notification solution based on a smartwatch that also promotes safe physical distancing in business, hospitality, or recreational facilities. Our proximity-based privacy-preserving contact tracing (P$^3$CT) leverages the Bluetooth Low Energy (BLE) technology for reliable proximity sensing, and an ambient signature protocol for preserving identity. Proximity sensing exploits the received signal strength (RSS) to detect the user's interaction and thus classifying them into low- or high-risk with respect to a patient diagnosed with an infectious disease. More precisely, a user is notified of their exposure based on their interactions, in terms of distance and time, with a patient. Our privacy-preserving protocol uses the ambient signatures to ensure that users' identities be anonymized. We demonstrate the feasibility of our proposed solution through extensive experimentation.
\end{abstract}


%
\IEEEpeerreviewmaketitle

\section{Introduction}
\label{sec:intro}
\noindent
\IEEEPARstart{M}{any} industries suspended their daily operations in correspondence to the government's effort in containing the COVID-19 pandemic. In view of the urgency to resuming the daily life routine, several countries have started to relax the restriction so that some industries can resume operation and have their employees back to normal activities. However, each industry is expected to implement some preventive measures to minimize the risk of further outbreaks. Among those preventive measures, such as temperature checks, face coverings, and frequent hand washing, contact tracing is deemed essential in monitoring the interaction between individuals and thus providing an immediate alert to all those who were exposed when someone is diagnosed with an infectious disease~\cite{ferretti2020quantifying},~\cite{eames2003contact}.

An efficient contact tracing approach needs to properly address the question of how to monitor the interactions between employees and customers and how to alert exposed individuals while preserving the anonymity of the patient.
While there are many smartphone-based contact tracing systems (e.g., Pan European Privacy-Preserving Proximity Tracing (PEPP-PT)~\cite{PEPP}, COVID-19 Watch~\cite{COVIDwatch}, Privacy-Preserving Automated Contact Tracing (PACT)~\cite{PACT}, etc.), these solutions might not be effective in a workplace because the employee does not necessarily carry with them the smartphone all the time due to the inherent nature of the activity.
Furthermore, many people might put the smartphone inside a pocket or backpack, which increases the difficulty in achieving reliable proximity sensing.
An effective and low-cost contact tracing solution that can be used by the employee without affecting their activity and at the same time providing line-of-sight (LOS) signals for more accurate proximity sensing is necessary.
Motivated by this limitation, this paper proposes a wearable contact tracing solution based on a low-cost smartwatch, namely proximity-based privacy-preserving contact tracing (P$^3$CT).

\begin{figure}[t!]
	\centering
	\includegraphics[width=0.95\columnwidth]{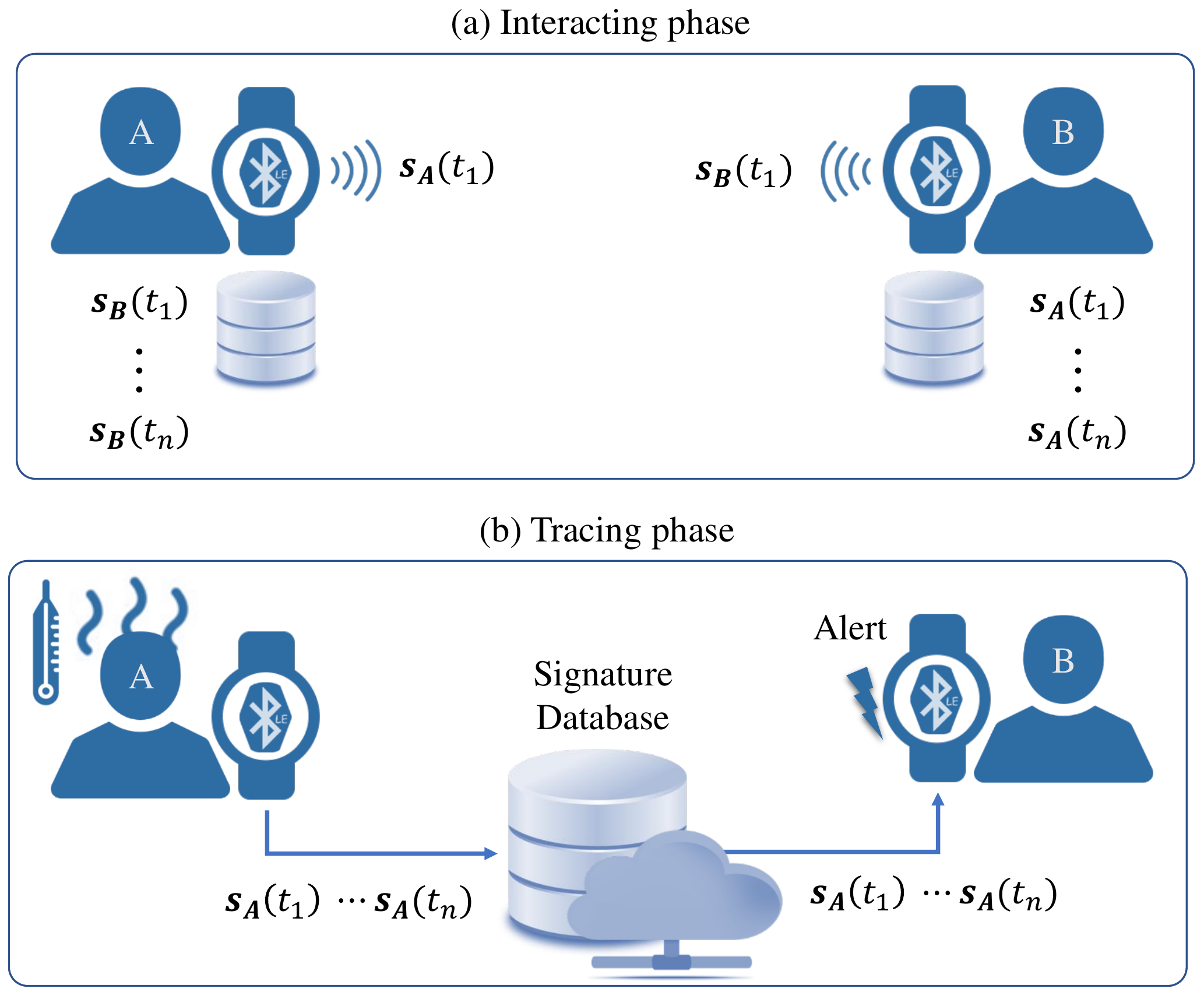}
	\caption{When employees A and B are in proximity to each other, their smartwatches will log the received BLE packet containing the ambient signature information into their watch's local storage. When employee A is diagnosed with an infectious disease, the watch will upload his/her own signatures to the signature database. All the other employees can download those signatures and compare them to a list of signatures they have observed in the past 14 days. An alert will be triggered when the downloaded signatures match one of the signatures on the list.}
	\label{fig:watchIllustration}
\end{figure}

First, we exploit the proximity sensing information computed from the received Bluetooth Low Energy (BLE) signals to monitor the interaction between employees~\cite{9000599}.
Second, we design a privacy-preserving protocol that encapsulates the BLE packet with an ambient signature packet rather than the employee's identity or location-related information.
The main framework describing the contact tracing based on BLE technology is shown in Fig.~\ref{fig:watchIllustration}.
Each smartwatch will broadcast a BLE packet periodically according to a system-defined interval. 
Rather than using the conventional two-way BLE communication channels (i.e., a secure channel for data exchange established through a series of pairing and handshaking processes), the smartwatch uses a non-connectable advertising channel, which was primarily used by beacon-based applications, to broadcast the packet. 
Hence, it is almost impossible for any malicious device to connect to the smartwatch to access sensitive information.

When two users are in proximity to each other, that is, when the smartwatches are within the broadcasting range, they can listen to the incoming packet and measure the received signal strength (RSS). 
The smartwatches will log the packet including the measured RSS value into their local storage, as shown in Fig.~\ref{fig:watchIllustration}(a).
The packet contains the ambient signature information observed by the individual smartwatch at particular timestamps. When an individual is diagnosed with an infectious disease, as shown in Fig.~\ref{fig:watchIllustration}(b), the smartwatch will upload the individual's own signatures generated for the past 14 days to the signature database.
All the other employees will download the infected signatures into their smartwatch for signature matching. 
In other words, the signature matching process takes place in the individual smartwatch rather than the cloud server. In this case, there is no way for others to know who has come into close contact with the infected person.
The smartwatch will automatically trigger an alert when it found a matched signature.
Based on the alert, the individual will be informed about the necessary actions, such as self-quarantine, testing, follow-up and monitoring process, so that further spread of this highly contagious disease can be prevented.

While it is relatively straightforward to develop such an application to the smartwatch for contact tracing purposes, it remains unclear how accurate is the proximity sensing information estimated through the RSS value and how the ambient signature information can help to prevent information leaks.
Furthermore, rather than a simple alert when there are matched signatures, it is more effective to tell the individual about their exposure risk level based on how close and how long their exposure was.
This is because the risk of an individual to be infected is low if they spent less than one second in close proximity to the infected employee compared to the individual who spent more than one hour in not so close proximity, yet still relatively near (i.e., the smartwatch still in the broadcasting range), to the infected employee.
Recognizing the above challenges, we carefully developed our proposed P$^3$CT that has the following contributions:
\begin{itemize}
    \item Accurate proximity sensing: this is the first work that provides a comprehensive investigation about the performance of proximity sensing based on the RSS values measured by a smartwatch worn by individuals. While RSS suffers severe attenuation due to the human body, our empirical analysis verified that we can achieve satisfactory performance with a carefully implemented machine learning method. Since the smartwatch is always worn on the human wrist, it is less challenging for the smartwatch to provide suitable LOS signals compared to smartphones.
    \item Low-cost device: Being a low-cost commercial off-the-shelf device that is equipped with essential BLE technology, the smartwatch has become an ideal solution for privacy-preserving contact tracing. The widely available software development support allows the workplace to provide quick yet reliable prototypes for testing prior to workplace reopening.
    \item Risk classification: we define the exposure's risk of a user based on the interaction duration and distance with the infectious individual. In contrast to most works that simply rely on the RSS value as an input feature to train a classification model, we explore other possible input features including the number of samples observed by the smartwatch, the maximum RSS, the minimum RSS, and the RSS range. Our experiment unveils the effects of selecting the right features on classification performance.
    \item Real-time notification and dataset: Our developed application can provide real-time notification when the physical distance is violated. Also, our experimental results were validated with a real-world dataset that was collected with a smartwatch worn on the human wrist.
\end{itemize}

The rest of the paper is organized as follows. 
Section~\ref{sec:ct} provides the background related to contact tracing and discusses its current development.
Section~\ref{sec:pppct} presents our proposed P$^3$CT.
Section~\ref{sec:riskClassification} describes the method to classify the risk level.
Section~\ref{sec:imp} discusses our experimental evaluations.
Section~\ref{sec:conclusions} concludes the paper with future works.

\section{Motivation to Contact Tracing}
\label{sec:ct}

Recognizing the urgency to have an effective contact tracing system, various digital-based solutions, either based on a smartphone or a smart wearable, have emerged lately.

\subsection{Major Phases of Contact Tracing}
During an epidemic of a highly contagious disease such as COVID-19, it is very likely for anyone to contract the virus when they interact with an infected individual in close proximity for a very long time. 
Contact tracing aims to trace down this group of people so that they can be aware of their exposure to the virus and take the necessary action as soon as possible. 
We can divide the contact tracing into two major phases:

\subsubsection{Interacting Phase}
The interacting phase keeps track of the daily contacts including distance and duration. A contact tracing system should be able to detect when any two persons are in proximity to each other at the same time keeping track of the duration they remain in close proximity.
An effective contact tracing system should be able to detect the proximity with high accuracy rather than seeking to estimate the exact distance, which is quite challenging considering the dynamic movement of humans.

\subsubsection{Tracing Phase}
When a person is diagnosed with the infectious disease, we need to trace down a list of people who have been in close contact with the infected person because they are more likely to get affected. 
If this group of people can be notified promptly, we reduce the chances for the virus to continue to spread to others.
However, many people are concerned about exposing their identity during the tracing phase.
Hence, a privacy-preserving contact tracing system should provide these two pieces of information without disclosing one's identity.

\subsection{Digital Tools for Contact Tracing}
The traditional contact tracing is conducted manually through in-person interviews and investigations.
Such a manual method based on subjective feedback (i.e., feedback from the infected individual) is unable to gather the precise interaction distance and duration. 
Furthermore, the investigator might acquire some sensitive information to identify those people who have come into close contact.
In contrast to the manual contact tracing, the digital tools for contact tracing can provide more precise information regarding the interaction distance and duration, at the same time preserving the individual's privacy.
To date, the digital-based contact tracing can be categorized into smartphone or smart wearable-based:

\subsubsection{Smartphone-Based Contact Tracing}
The pervasiveness of smartphones has made smartphones the most popular choice when comes into the digital-based contact tracing system. 
The rich sensing features embedded in the smartphone provides a better estimation of interaction distance and duration~\cite{chen, werner,liu}.
For example, many works leverage location sensing~\cite{skCorona}  and proximity sensing~\cite{PEPP} to keep track of the interaction between any two individuals.
There are also works exploiting the heterogeneous sensing features in a smartphone to improve the distance estimation~\cite{nguyen2020epidemic}.
However, most of these works fail to consider the location of the smartphone during the interacting phase. 
While people might carry their smartphone with them during grocery shopping, the smartphone will be inside a pocket or a purse most of the time. Hence, the distance estimation is more complex and can be highly inaccurate for a contact tracing application.
At the same time, people might not carry the smartphone with them all the time while working.

\subsubsection{Wearable-Based Contact Tracing}
Considered the inconsistency of smartphones, some companies have started to exploit the smart wearable approach to contact tracing.
The goal is to resume the working routine with less distraction.
For example, EasyBand~\cite{tripathy2020easyband} presents a wearable solution to auto contact tracing while encouraging safe social distancing practice during interaction.
However, EasyBand uses a centralized server for contact tracing, in which all the users' data is uploaded to the cloud through TCP/IP connection.
Such a centralized approach is not scalable as all the computations to find the close contact for all the workers are performed within the server. 
Furthermore, there is a high possibility of information leak if the server is compromised.
Our proposed P$^3$CT, on the other hand, provides a privacy-preserving contact tracing by keeping no individual information on the cloud server.

\subsection{Current Development in Contact Tracing}
Recognizing the importance of contact tracing in resuming the normal lifestyle while preventing the further virus outbreak, both government and academia have devoted efforts in developing a more effective contact tracing solution to fight against COVID-19.

\subsubsection{National-Level Efforts}
China, South Korea, and Singapore are among the first countries enforcing digital tools for contact tracing. 
China leverages its existing surveillance strategy to implement a close contact detector based on QR codes technology~\cite{chinaCorona}. 
South Korea leverages the location data (i.e., the GPS data) from the smartphone to detect the distance of the users and push a notification containing personal details of the infected individuals to the nearby users~\cite{skCorona}.
Singapore developed the TraceTogether application based on BLE signals on the smartphone to detect the proximity between any two individuals~\cite{sgCorona}.
While the methods applied by China and South Korea might be less strict on user's privacy, Singapore adopted a more privacy-preserving approach by only tracking the proximity between users without explicit location information.

\subsubsection{Academia-Level Efforts} 
There have been a number of initiatives from industry and academia researchers in delivering an effective contact tracing solution while preserving user privacy~\cite{shukla2020privacy},~\cite{bell2020tracesecure}.
For example, Pan European Privacy-Preserving Proximity Tracing (PEPP-PT) detects the proximity based on the broadcast BLE packet containing a full anonymous ID~\cite{PEPP}.
COVID-19 Watch provides automatic alert the user when they are in contact with the infected individual~\cite{COVIDwatch}.
The Privacy-Preserving Automated Contact Tracing (PACT) exploits the BLE signals in combination with secure encryption to detect possible contacts while protecting users' privacy~\cite{PACT}.

Most of these initiatives assume that the BLE signals will work for proximity detection while there are no works providing a comprehensive study of the accuracy of using BLE signals for proximity sensing. To bridge the gap, this paper presents extensive experiments to validate the feasibility of using BLE signals for proximity detection.

\section{Proposed Proximity-Based Privacy-Preserving Contact Tracing}
\label{sec:pppct}
Our proposed P$^3$CT leverages the BLE technology available on the smartwatch for proximity sensing. 
To achieve privacy-preserving contact tracing, we adopt the following signature protocol to define the BLE advertising packet.


\subsection{Proximity Sensing with BLE Technology}
\label{ss:ble}
As a popular short-range communication over the 2.4~GHz ISM band~\cite{gomez2012overview}, BLE is readily available in many smart devices including smartwatches, earphones, smart thermostats, etc.~\cite{7000963, 7366936}.
BLE communicates through either non-connectable advertising or connectable advertising~\cite{8011489}.
The latter advertising mode allows another device to request a secure connection through handshaking.
Our proposed P$^3$CT  uses the non-connectable advertising mode, which rejects any incoming connection requests~\cite{8242361}, as the main feature for exposure notification purposes. 
The non-connectable advertising mode allows the smartwatch to broadcast a short advertising packet periodically according to the system-defined advertising interval, $T_a$. Upon receiving the packet, the smartwatch can measure the RSS and use it to estimate the proximity.
More precisely, the RSS is inversely proportional to the square of the distance according to the inverse square law \cite{8395148, 8423010}:
\begin{equation}
P_r \propto \frac{1}{d^n}
\end{equation}
where $P_r$ is the signal strength in dBm, $d$ is the distance between any two smartwatches, and $n$ is the path loss exponent. 

\begin{figure}[t!]
	\centering
	\includegraphics[width=0.95\columnwidth]{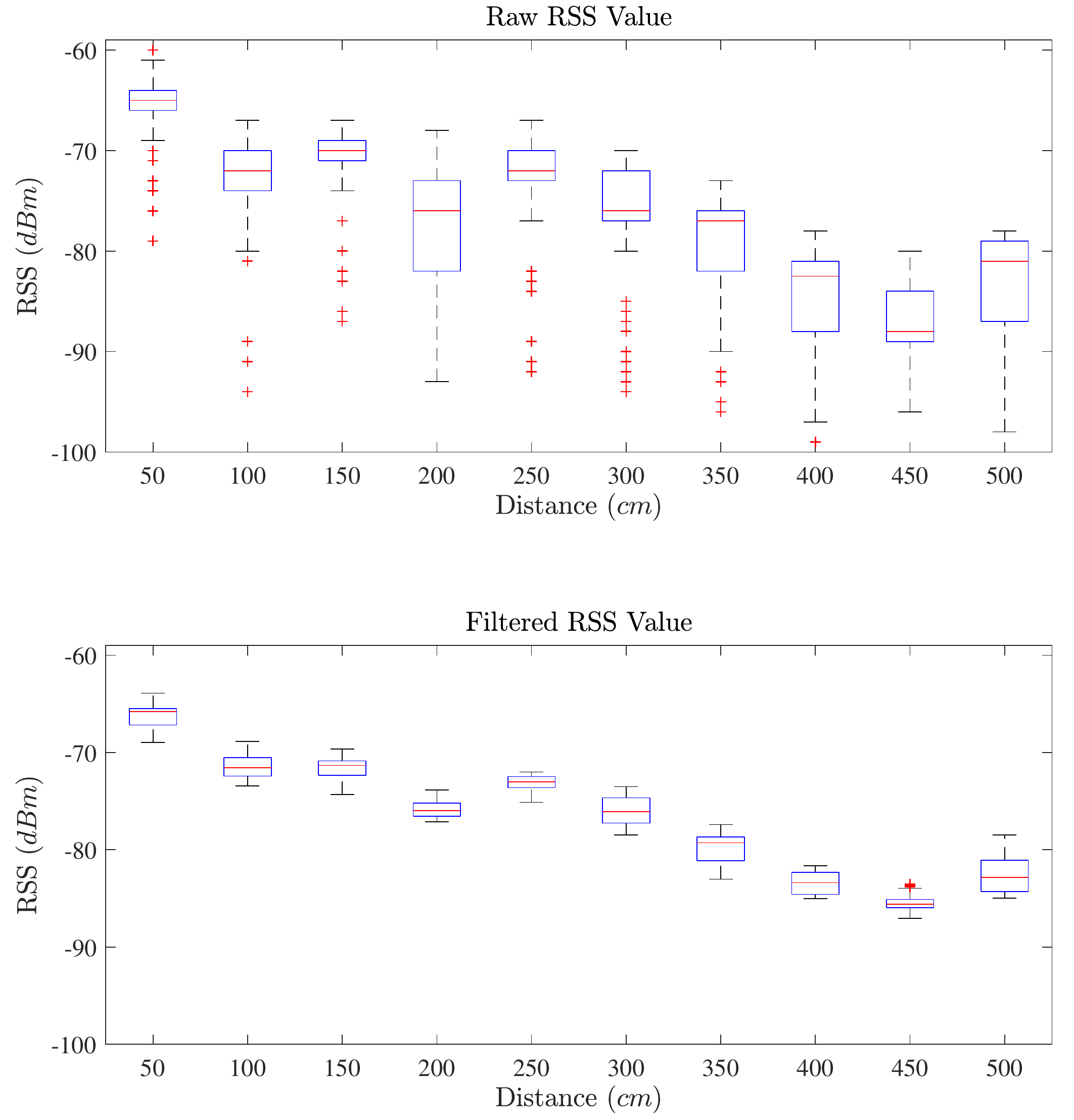}
	\caption{Variation of RSS over distance without and with filtering.}
	\label{fig:rawFiltered_dist}
\end{figure}

Even though the RSS-distance relationship holds for the signal in the free space, the RSS values suffer a great variation in practical environments owing to the multipath~\cite{7174982} and body shadowing effects~\cite{6856188, 8588347}.
We can minimize the signal variation by applying signal filtering methods, such as moving average.
As shown in Fig.~\ref{fig:rawFiltered_dist}, the RSS values at each distance are more distinct and with less variation when a moving average is applied  as compared to the raw RSS data.
While we can set a cut-off threshold, for example, any value greater than $-75$~dBm as being in close proximity, such a thresholding approach will result  in high false-negative  with raw RSS value (i.e. the system will not record the contact as close proximity) and high false-positive with filtered RSS value (i.e. the system will record the contact in close proximity while it is not).
Rather than using a thresholding approach, we exploit machine learning methods to proximity sensing and further classify the sensing output into high-risk and low-risk.

\subsection{Privacy-Preserving Signature Protocol}
\label{sec:signature}
We propose a signature protocol that constructs a signature vector that fits into the length-constrained advertising packet (i.e., the available payload is only 31~bytes).
Specifically, each smartwatch is configured to execute the following functions:
\begin{enumerate}[label=\roman*.]
	\item Signature generation: The smartphone scans for the ambient environmental features. These features are selectively processed to generate a unique signature that fits into the 31 bytes advertising payload. The signature is  updated every few minutes.
	\item Signature broadcasting: The smartphone broadcasts the advertising packet containing the unique signature periodically according to the advertising interval $T_a$. The packet is broadcasted through the non-connectable advertising channels.
	\item Signatures observation: The smartphone scans the three advertising channels to listen for the advertising packet broadcast by the neighboring smartphones. The scanning is performed in between the broadcasting events.
\end{enumerate}
 
The signature is a 31-dimensional transformed vector containing the ambient environmental features.
Upon the generation of the signature, the smartwatch will encapsulate the signature information into its advertising packet and broadcast the packet through the non-connectable advertising channels.
The nearby smartwatches can see the packet when they scan on those advertising channels where the packet is transmitted.

\begin{figure}[t!]
	\centering
	\includegraphics[width=1\columnwidth]{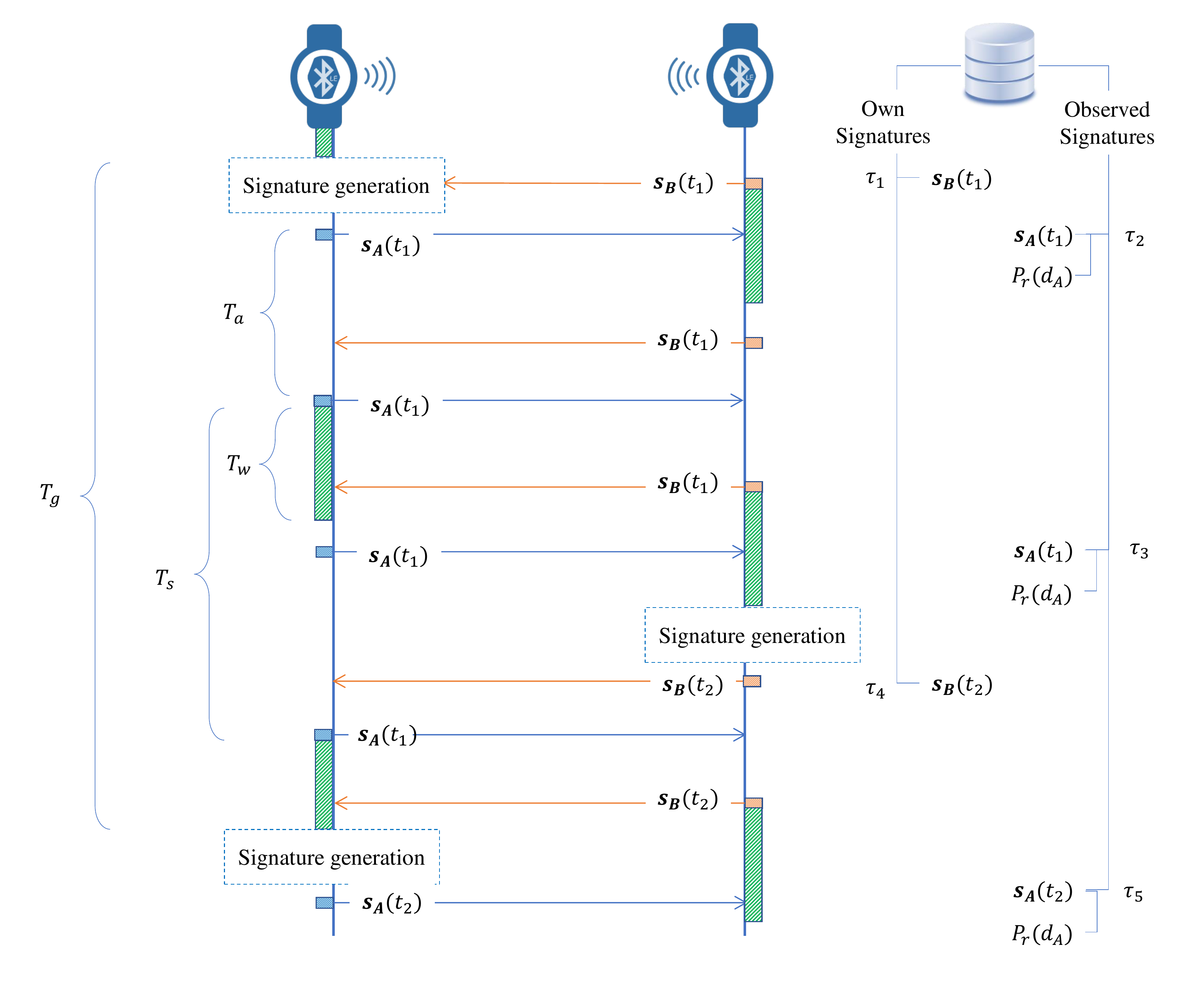}
	\caption{Timing diagram for the advertising, scanning and signature generation activities. All the generated and observed signatures will be logged in the local database, together with a timestamp $\tau$.}
	\label{fig:timeDia}
\end{figure}

The timing diagram for the advertising, scanning, and signature generation activities, in which each activity is triggered periodically according to its interval, i.e., generation interval $T_g$, advertising interval $T_a$, and scanning interval $T_s$, is shown in Fig.~\ref{fig:timeDia}. 
Given $T_s$, the smartwatch will only stay active to listen for the incoming packet for a duration defined by the scanning window $T_w$.
While it is possible to use a continuous scanning (i.e., by setting $T_w = T_s$) to increase the packet receiving rate, such a scanning approach has an adverse effect on the energy consumption.

\section{Risk Classification}
\label{sec:riskClassification}
Besides using proximity sensing to detect the approximate interaction distance between any two individuals, we also consider the interaction time when labeling the individual into low or high risk.

\subsection{Proximity Sensing}
Proximity sensing has been employed in many scenarios, for example, to identify the user proximity to museum collection~\cite{9001059}, gallery art pieces~\cite{8019467}, etc. Some works study the proximity detection in a dense environment~\cite{8059756}, or proximity accuracy with filtering technique~\cite{8956048}. However, most of these works study the proximity detection between a human and an object with an attached BLE beacon~\cite{8705339}. So far, there is no work studying the proximity sensing between devices carried by two humans. While estimating the distance can help to check if the user maintains a safe physical distancing, an exact 2~m distance should not be a rigid requirement in classifying the risk of a user.
Rather, we are more interested to know the proximity between any two workers, and how long they remain in proximity. Then, we can forward these pieces of information to the  epidemiologists and they can decide to classify a contact as high or low risk.
BLE is an excellent technology for the above purpose since BLE is a short-range communication that can only be heard when two smartwatches are in the communication range of each other.
Upon receiving the advertising packet, the smartwatch can measure the RSS and thus estimate its proximity to the nearby smartwatch.
We classify the proximity into two classes, i.e., far and close. 
We define close proximity when the distance between any two smartwatches is less than $2$ m, and any distance greater than $2$ m but less than the broadcasting range is considered far. 
In other words, the two smartwatches are not in proximity if they are outside the broadcasting range of each other.

\begin{figure}[t!]
	\centering
	\includegraphics[width=0.95\columnwidth]{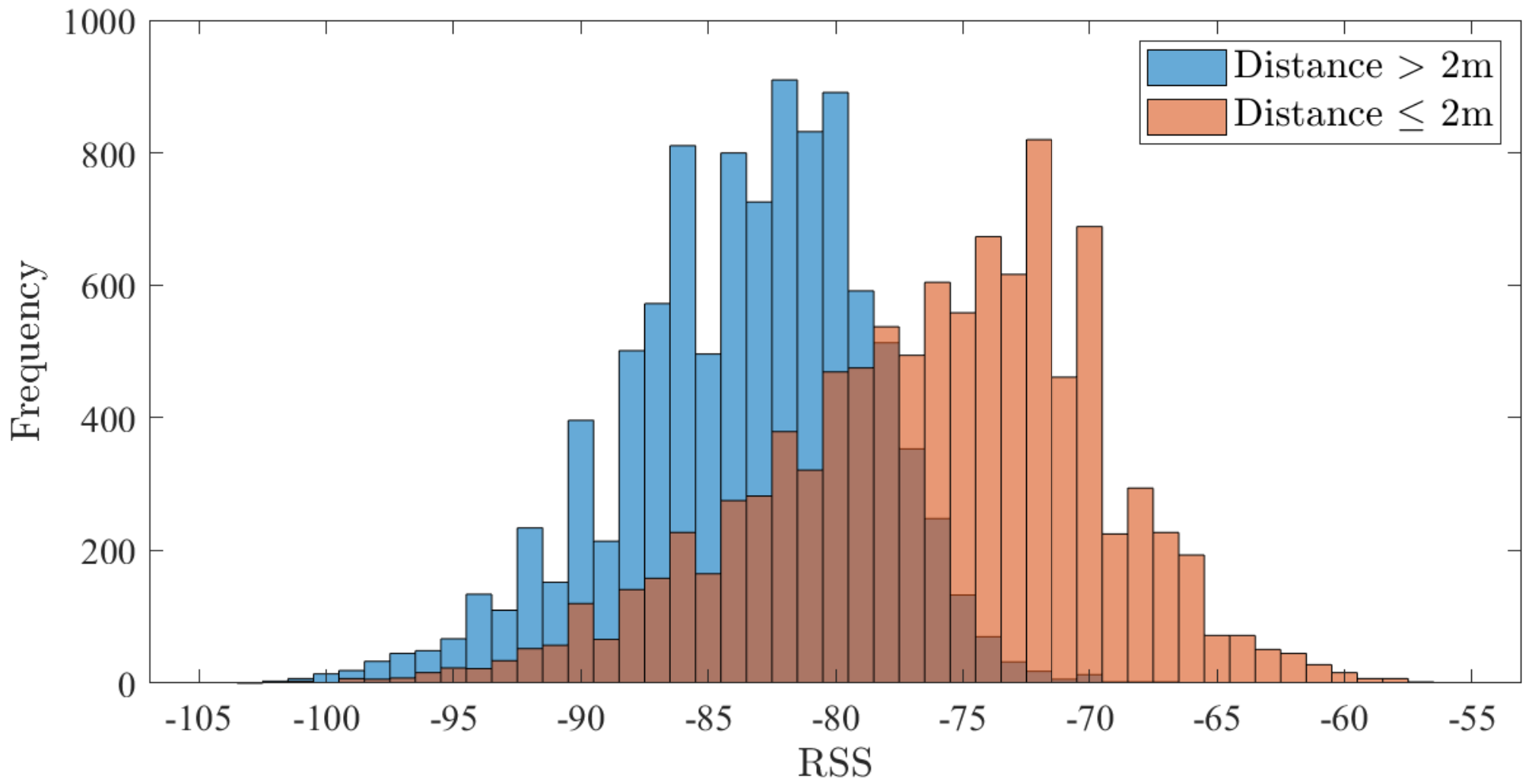}
	\caption{RSS distributions for two types of proximity: far (blue color bars) and close (orange color bars).}
	\label{fig:cutDownRSSdistribution}
\end{figure}

The RSS distributions for far and close proximity is shown in Fig.~\ref{fig:cutDownRSSdistribution}.
It is clear that there would be errors if proximity were decided by simply setting an RSS threshold. For example, if we set anything above $-80$~dBm as close proximity, chances are some values greater than $-80$~dBm are from the smartwatch located at a distance greater than $2$ m.
Hence, it is unreliable to identify the risk of an individual simply based on the proximity. 
At the same time, some individuals might come in very close proximity when they pass by each other.
Hence, we also consider the interaction time when we want to identify the risk of an individual.

\subsection{Hypothesis to Risk Classification}
\begin{figure}[t!]
	\centering
	\includegraphics[width=0.95\columnwidth]{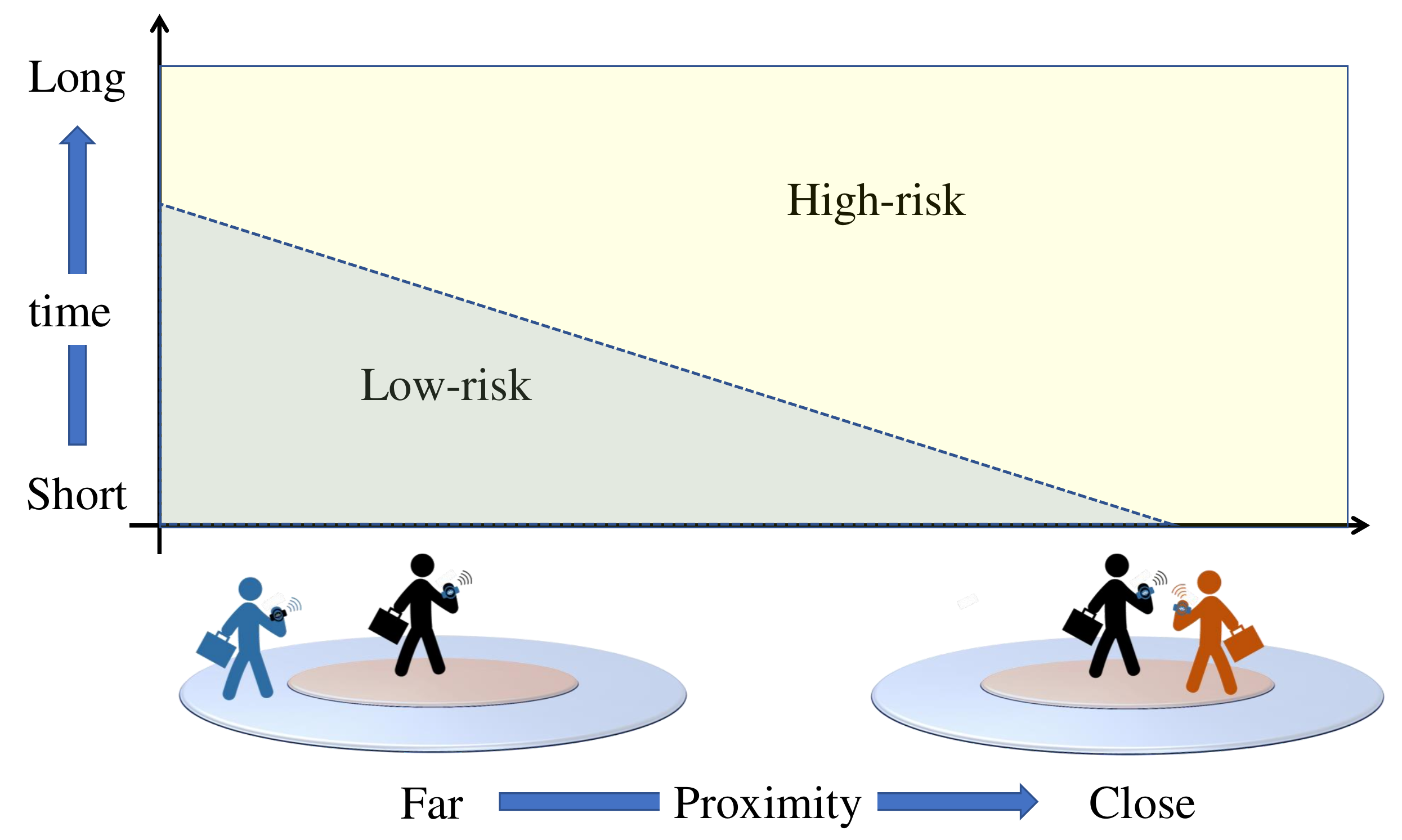}
	\caption{The individual is considered in the high-risk group of being infected when he/she is in close proximity with the infected individual or has spent a very long time with the infected individual within a confined space even though the proximity between them might be high.}
	\label{fig:riskHyp}
\end{figure}

While it is more likely to be infected when the individual is in close proximity to the infected person, the risk of getting infected is relatively low if the individual only spends less than 1 second in such close proximity.
On the other hand, the risk of getting infected can be high if the individual spends a very long time together with the infected person even if they are not in very close proximity. While the exact definition should be left to the epidemiologists, we can provide them with the necessary information which includes the distance and the time of the interaction.
The possible risk of getting infected with respect to the proximity and interaction time between the individual and the infected person is shown in Fig.~\ref{fig:riskHyp}.

The problem of classifying the risk of a potential contact can be modeled as a binary hypothesis test.
Let $\mathbf{x}$ be an $m$-dimensional feature vector and consider a risk mapping function $R:(\mathbf{x}) \longrightarrow \{ +1, -1 \}$, where $+1$ indicates high-risk and $-1$ low-risk, then we have the following three hypotheses:
\begin{equation}
\begin{aligned}
&H_0: R(\mathbf{x}) = 0 \\
&H_+: R(\mathbf{x}) = +1  \\
&H_-: R(\mathbf{x}) = -1
\end{aligned}
\end{equation}
where $H_+$ denote the hypothesis that the user belongs to the high-risk ($+1$) group, $H_-$ the hypothesis that the user belongs to the low-risk ($-1$), and $H_0$ the null hypothesis.
Specifically, the null hypothesis happens when the user is risk-free, i.e., the user is outside the communication range of the infected person.
Obviously, miss detection is undesirable because the user might be at risk but the system considers the user safe. False-negative misclassified the high-risk user to low-risk, this may give a wrong impression to the user that the probability for them of getting infected is low, but actually, the probability could be high. 
While false-positive is a bit more conservative by misclassifying the low-risk user to high-risk, it is a relatively safer outcome than miss detection and false-negative.

\begin{figure}
	\centering
	\includegraphics[width=1\columnwidth]{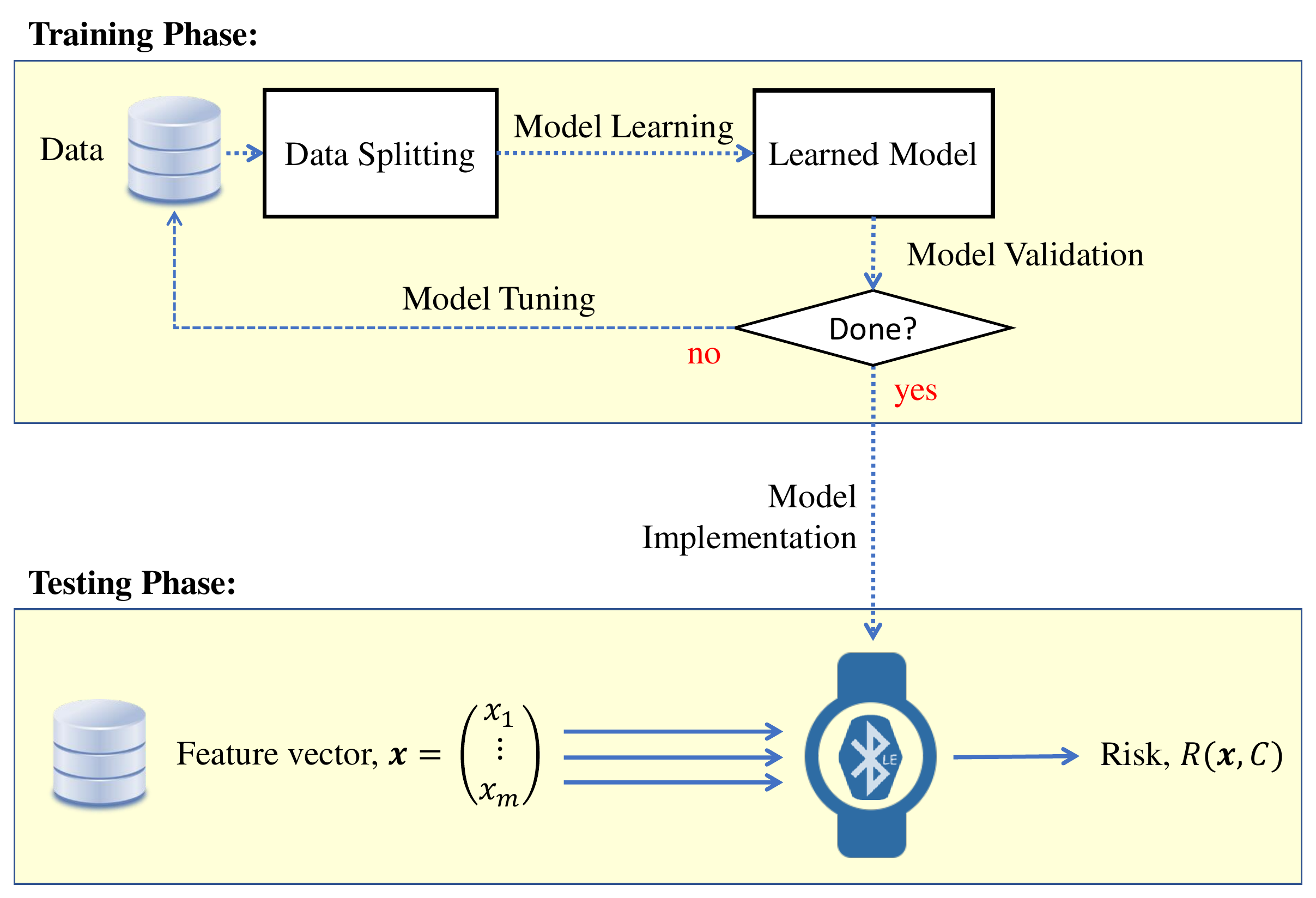}
	\caption{Classification model can be learned, validated and fine-tuned using the collected data. The final model can be loaded to the smartwatch to classify the risk of a user given a set of feature vectors stored in the smartwatch for the previous 14 days.}
	\label{fig:classificationTrainingFlow}
\end{figure} 
\subsection{Classification Models}
We can apply supervised machine learning methods to train a classification model. 
However, supervised methods required a set of labeled data, which is not readily available in the context of smartwatches.
In contrast to the abundant and open-accessible sources of text-based (e.g., WikiLens, BookCrossing, etc.) or image-based (e.g., MNIST, imageNet, etc.)  dataset, there are not dataset about the BLE signals received by the smartwatch.
To address this problem, we developed an application on the smartwatch to collect the BLE data.
Given the collected data, we can train a classification model, as shown in Fig.~\ref{fig:classificationTrainingFlow}.

During the training phase, the data is split into training and validation set before feeding the data for model learning.
The objective is to learn a set of weights that fits the hypothesis function $R(\mathbf{x}, C)$ defined by the corresponding classification model $C$.
Validation is performed to evaluate the learned model as well as preventing the model from overfitting.
If necessary, model fine-tuning can be performed to improve the classification performance.
Mathematically, the learning process aims to fit the risk mapping function $R:(\mathbf{x}) \longrightarrow y$ given a set of $n$ training samples $\{(\mathbf{x}_1, y_1), \dots, (\mathbf{x}_n, y_n)\}$, where $\mathbf{x} = (x_1, \dots, x_m)^T$ is an $m$-dimensional feature vector and $y = \{ +1, -1 \}$ is the classification output indicating the risk of a user.

In this paper, we exploit four types of classifications: decision tree (DT), linear discriminant analysis (LDA), na\"{i}ve Bayes (NB) and k-nearest neighbors (kNN).

\subsubsection{DT}
Top-down approach is the commonly used method to learn a classification tree. More precisely, DT starts by choosing a feature from the feature vector that provides the best splitting in connection to the target risk label, and then repeats the same splitting procedures for each separate branch until it reaches a final decision.
Let $\theta = (x, \gamma)$ be the splitting rule given feature $x$ and threshold $\gamma$, we can split $n$ samples of training data $\mathcal{T}$ into two subsets, i.e., 
\begin{equation}
\begin{aligned}
& \mathcal{T}_r(\theta) = (\mathbf{x}, y) | x \leq \gamma \\
& \mathcal{T}_l(\theta) = \mathcal{T} \setminus \mathcal{T}_r(\theta)
\end{aligned}
\end{equation}
where $\mathcal{T}_r$ and $\mathcal{T}_l$ are the resultant subsets representing the data for right and left branches, respectively.
The commonly measure used to govern the splitting rule is the Gini impurity $G(\cdot)$, which tells how likely the model will produce a misclassification if the model predicts the labels based on the labels distribution from a randomly chosen feature.
Mathematically, the Gini impurity can be computed as follows:
\begin{equation}
G(\mathcal{T}, \theta) = \frac{n_l}{n} \mathbb{H}(\mathcal{T}_l(\theta)) + \frac{n_r}{n} \mathbb{H}(\mathcal{T}_r(\theta))
\end{equation}
where $n_l$ and $n_r$ are the number of training samples for each subset, and $\mathbb{H}(\cdot)$ is the entropy function, i.e.,
\begin{equation}
\begin{aligned}
\mathbb{H}(\mathbf{x}) &= \sum_{y = \{ +1, -1 \}} p_y (1 - p_y) \\
&= - \sum_{y = \{ +1, -1 \}} p_y \log(p_y)
\end{aligned}
\end{equation}
and $p_y$ denotes the probability of correct classification. Suppose that $\mathbb{I} = \{1, 0\}$ be the indication function and $\tilde{y}$ be the predicted output, then we have
\begin{equation}
p_y = \frac{1}{n} \sum_{\forall x \in \mathbf{x}} \mathbb{I}(\tilde{y} = y)
\end{equation}
The objective of DT is to find the parameters that produce the best splitting rule, i.e.,
\begin{equation}
\theta^* = \argmin G(\mathcal{T}, \theta)
\end{equation}
\subsubsection{LDA}
Assuming that the covariance for each class is the same, LDA learns a classifier by fitting a Gaussian density to each class.
Let $\mathbb{P}(\mathbf{x}|\tilde{y} = y)$ be the conditional distribution for each class $y = \{ +1, -1 \}$, by applying Bayes' rule, we obtain
\begin{equation}
\mathbb{P}(\tilde{y} = y | \mathbf{x}) = \frac{\mathbb{P}(\mathbf{x}|\tilde{y} = y) \mathbb{P}(\tilde{y} = y)}{ \sum_{y = \{ +1, -1 \}}\mathbb{P}(\mathbf{x} |y)\mathbb{P}(y)}
\end{equation}
Then, the class (i.e., the risk) can be determined by selecting the output with the highest posterior probability.

\subsubsection{NB}
Following a na\"{i}ve assumption that each feature is conditionally independent, we can apply the Bayes' theorem to learn a classification model. 
By simplifying $\mathbb{P}(x|y, \forall x \in \mathbf{x})$ to $\mathbb{P}(x|y)$, we have
\begin{equation}
\mathbb{P}(y|\forall x \in \mathbf{x}) = \frac{\mathbb{P}(y) \prod_{i=1}^{m}\mathbb{P}(x|y) }{\mathbb{P}(\mathbf{x})}.
\end{equation}
Since $\mathbb{P}(y|\forall x \in \mathbf{x})$ is proportional to $\mathbb{P}(y) \prod_{i=1}^{m}\mathbb{P}(x|y)$, then we can use maximum a posteriori (MAP) to estimate the probability for each class $\mathbb{P}(y)$ and the conditional probability for each class given the feature $\mathbb{P}(x|y)$. The output risk can then be predicted based on the following rule:
\begin{equation}
\tilde{y} = \argmax_{y = \{ +1, -1 \}} \mathbb{P}(y) \prod_{i=1}^{m}\mathbb{P}(x|y)
\end{equation}

\subsubsection{kNN}
The goal of kNN is to maximize the probability of correct classification. Let $p_i$ indicate the probability that a training sample $i$ is classified correctly, according to the stochastic nearest neighbors' rule, we have:
\begin{equation}
p_i = \sum_{j \in \mathcal{T}_i} p_{ij}
\end{equation}
where $\mathcal{T}_i$ is a subset of data belonging to the same class as training sample.
Given $p_i$, the goal of kNN can be defined as follows:
\begin{equation}
\argmax_{y = \{ +1, -1 \}} \sum_{i=1}^{n} p_i
\end{equation}

Note that all the classifiers described above can be further extended by assuming different distribution functions. One of the possible future work is to calibrate the classifier based on the prior empirical distribution knowledge about a certain environment. More precisely, different environments might produce different distributions, and if we can acquire this information, it could help to better calibrate the classifier and thus improve the classification performance.

\section{Experiments and Evaluations}
\label{sec:imp}

We consolidated the collected data from both smartwatches before dividing them into training and testing datasets.
Then, we evaluate the experimental results obtained from different classifiers.

\subsection{Experimental Setup with Smartwatch}
\begin{figure}[t!]
	\centering
	\includegraphics[width=0.9\columnwidth]{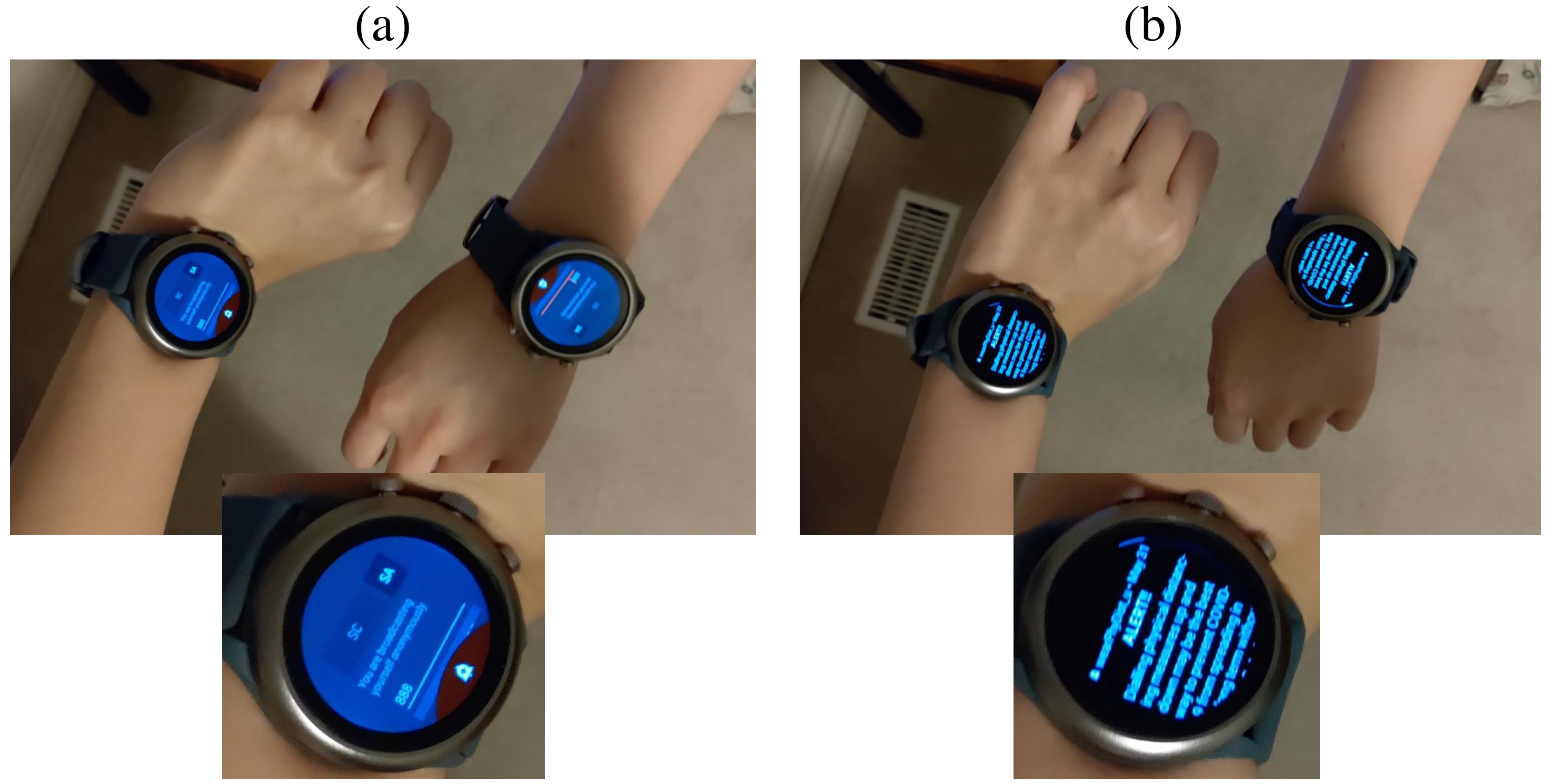}
	\caption{When any two persons come close to each other, (a) the smartwatch vibrates with an alert, and (b) it also push a notification to remind the users to practice safe physical distancing.}
	\label{fig:watchAlert}
\end{figure}

For the experiment, we used Fossil Sport, a smartwatch based on  Google's Wear OS 2.17.
The smartwatch is powered by a Qualcomm Snapdragon Wear 3100 processor and has an internal memory of up to 1~GB.
The 8~GB internal storage is sufficient to store the generated and observed signatures for at least 14~days. 
The small form factor (i.e., 1.28 in AMOLED screen with 44~mm case size and 12~mm case thickness) makes the smartwatch an ideal candidate for contact tracing in the workplace. 
As shown in Fig.~\ref{fig:watchAlert}, the smartwatch can trigger the alert automatically when any two smartwatches are in close proximity to each other.

\begin{figure}
	\centering
	\includegraphics[width=0.9\columnwidth]{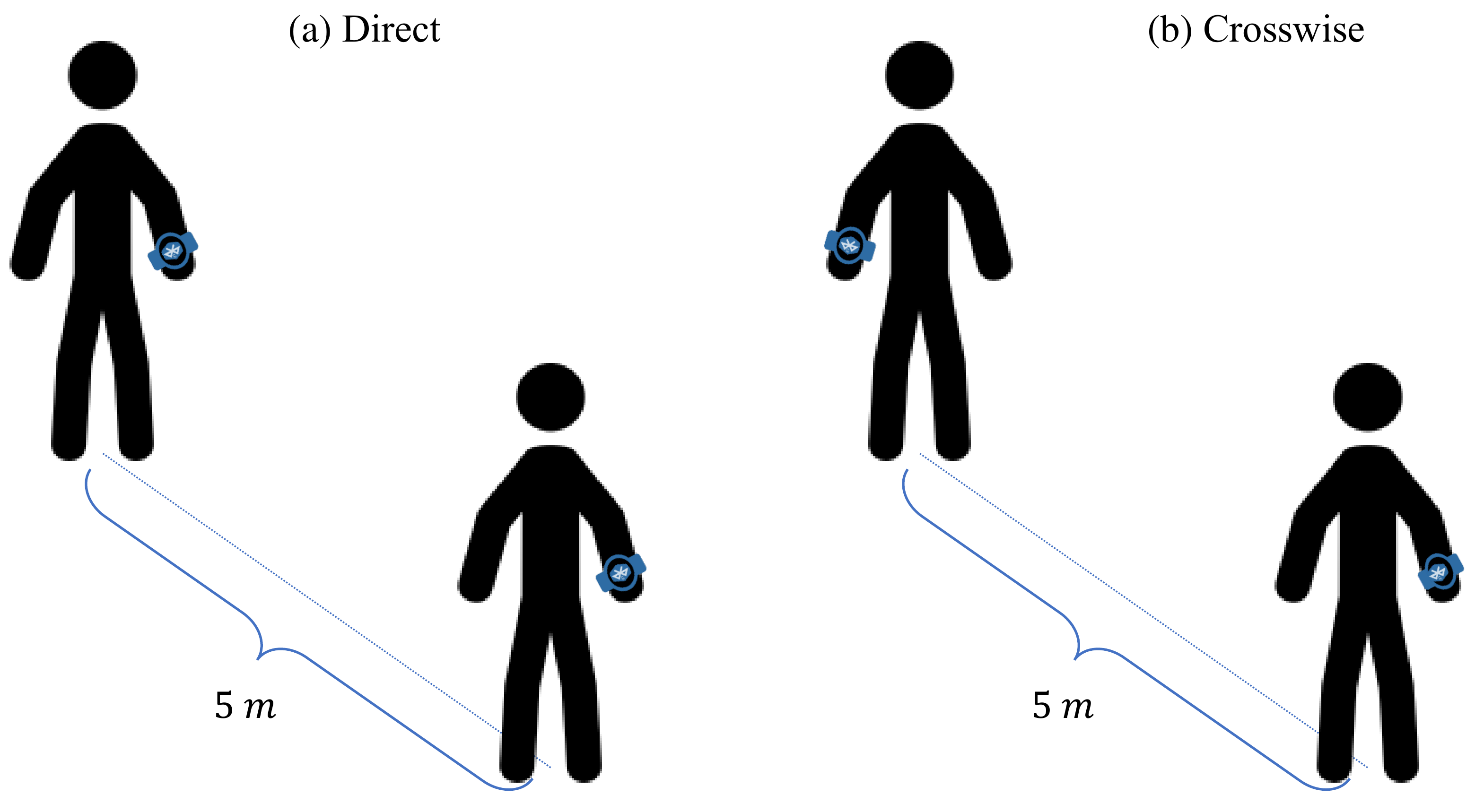}
	\caption{Four combinations of smartwatch on hand's position have been tested, i.e., left hand to right hand (LR), right hand to left hand (RL), left hand to left hand (LL) and right hand to right hand (RR). These four combinations can be classified into two categories: (a) direct line (LR and RL) and (b) crosswise line (LL and RR).}
	\label{fig:experimentSetup}
\end{figure}

We programmed the smartwatch to broadcast the advertising packet in the background. For experimental purposes, we also programmed the application to log all the advertising packet it received at every distance.
In particular, the following information will be logged: the ground truth  distance, name of the smartphone, MAC address of BLE chipset, the packet payload, RSS values, time elapsed, and timestamp.
The time elapsed indicates the time difference between the previous broadcast packet and the current broadcast packet, whereas the timestamp is the exact time when the smartphone received the packet.
We performed the experiment by asking two volunteers to stand at a certain distance from each other, from 0.5~m up to 5~m, as illustrated in Fig.~\ref{fig:experimentSetup}.
A measuring tape is used as a reference to the ground truth distance.

\begin{table}[t!]
	\caption{Total data from each combination}
	\label{table:dataPoints}
	\centering  
	\begin{tabular}{cx{3cm}rx{3cm}}  
		\toprule
		\cmidrule{1-2}
		Combination     &  Total Data Points  \\					
		\midrule
		RR 		& 8168 \\
		LL      & 7874 \\
		RL   	& 13117 \\
		LR   	& 8485 \\
		\bottomrule
	\end{tabular}
\end{table}

We first performed the experiment by asking volunteer A to wear the smartwatch on her left hand, and volunteer B on her right hand (i.e., left to right (LR)). After that, we repeated the same experiment with right hand to left hand (RL), left hand to left hand (LL), and right hand to right hand (RR).
Since LR and RL constitute a direct line between two smartwatches and LL and RR constitute a crosswise line, we categorize these four hand-combinations into two groups: a) direct line, and b) crosswise line.
All the measurement data is saved into a ``comma-separated values" (.csv) file format and exported to Matlab for training and testing.

\begin{table*}[t!]
	\caption{The performance of each classifier for the direct dataset}
	\label{table:performanceS}
	\centering  
	\begin{tabular}{l|cc|cc|cc|cc}  
		\toprule
		\cmidrule{1-9}
		\multirow{2}{*}{\textbf{Classifier}}     &	
		\multicolumn{2}{c|}{\textbf{Precision}} & \multicolumn{2}{c|}{\textbf{Recall}}  & \multicolumn{2}{c|}{\textbf{F1-score}} & \multicolumn{2}{c}{\textbf{Accuracy}}\\
		&         Mean  & 95\% CI & Mean  & 95\% CI  &Mean  & 95\% CI & Mean  & 95\% CI\\  					
		\midrule
		DT 		& 0.9670 &(0.9653, 0.9686) & 0.9330 &(0.9305, 0.9353)  & 0.9497 &(0.9483, 0.9511)  & 0.9416 &(0.9400, 0.9431)  \\
		LDA     & 0.8600 &(0.8572, 0.8628) & 0.8439 &(0.8406, 0.8472)  & 0.8518 &(0.8497, 0.8541)  & 0.8293 &(0.8268, 0.8318) \\
		NB   	& 0.9562 &(0.9548, 0.9577) & 0.9127 &(0.9099, 0.9152)  & 0.9339 &(0.9324, 0.9355)  & 0.9228 &(0.9212, 0.9247) \\
		kNN   	& 0.8853 &(0.8828, 0.8882) & 0.9170 &(0.9145, 0.9194)  & 0.9009 &(0.8994, 0.9028)  & 0.8889 &(0.8872, 0.8909) \\
		\bottomrule
	\end{tabular}
\end{table*}
\begin{table*}[t!]
	\caption{The performance of each classifier for the crosswise dataset}
	\label{table:performanceO}
	\centering  
	\begin{tabular}{l|cc|cc|cc|cc}  
		\toprule
		\cmidrule{1-9}
		\multirow{2}{*}{\textbf{Classifier}}     &	
		\multicolumn{2}{c|}{\textbf{Precision}} & \multicolumn{2}{c|}{\textbf{Recall}}  & \multicolumn{2}{c|}{\textbf{F1-score}} & \multicolumn{2}{c}{\textbf{Accuracy}}\\
		&         Mean  & 95\% CI & Mean  & 95\% CI  &Mean  & 95\% CI & Mean  & 95\% CI\\  
		\midrule
		DT 		& 0.9176 &(0.9140, 0.9208) & 0.9133 &(0.9104, 0.9157)  & 0.9154 &(0.9133, 0.9175)  & 0.9059 &(0.9037, 0.9083)  \\
		LDA     & 0.7959 &(0.7912, 0.8008) & 0.8255 &(0.8224, 0.8292)  & 0.8104 &(0.8070, 0.8137)  & 0.7934 &(0.7903, 0.7966) \\
		NB   	& 0.9247 &(0.9212, 0.9272) & 0.8622 &(0.8590, 0.8650)  & 0.8924 &(0.8902, 0.8948)  & 0.8763 &(0.8739, 0.8788) \\
		kNN   	& 0.9392 &(0.9354, 0.9422) & 0.8270 &(0.8238, 0.8319)  & 0.8795 &(0.8767, 0.8820)  & 0.8573 &(0.8541, 0.8600) \\
		\bottomrule
	\end{tabular}
\end{table*} 
\subsection{Data Preparation and Processing}
In total, we have collected 37,644 data points from all the four combinations, as shown in Table~\ref{table:dataPoints}. 
We consolidated the data from RR and LL into a single dataset (i.e., the crosswise dataset) and then apply an 80\%--20\% splitting rule to split the data into training and testing set.
Similarly, we applied the same splitting rule to the consolidated data from RL and LR (i.e., the direct dataset).
For each training and testing set, the first four columns indicate the input features and the last column is the target label (i.e., the risk).
These four input features include the number of samples observed by the smartwatch, mean RSS, maximum RSS, minimum RSS, and the RSS range (i.e., maximum RSS $-$ minimum RSS).
Note that the number of samples observed by the smartwatch tells how long the smartwatch being in proximity to each other. 
The final training and testing data for both sets are shared openly in our GitHub repository~\cite{git}.

\subsection{Evaluation Metrics}
We used four metrics (i.e., precision ($p$), recall ($r$), F1-score ($f_1$) and accuracy ($a$)) to evaluate the performance of the classifier. 
Let $T^+$, $T^-$, $F^+$ and $F^-$ denote the true-positive, true-negative, false-positive and false-negative, respectively, then the above four metrics can be computed as follows:
\begin{equation}
p = \frac{T^+}{T^+ + F^+}
\end{equation}
\begin{equation}
r = \frac{T^+}{T^+ + F^-}
\end{equation}
\begin{equation}
f_1 = 2\frac{rp}{r+p}
\end{equation}
\begin{equation}
a = \frac{T^+ + T^-}{T^+ + T^- + F^+ + F^-}
\end{equation}

Precision tells how many are actually in the high-risk of all the classifier predicted as positive. 
In other words, high precision indicates the classifier produces low false-positive, which means the classifier is capable of avoiding create unnecessary tension and anxiety to the people.
Recall, on the other hand, tells how many we predicted as high-risk are in fact high-risk of being infected. 
In contrast to the accuracy that considers the number of correctly classified true-positives and true-negatives, F1-score considers the balance of precision and recall. 
F1-score is a useful metric when false-negatives and false-positives are important factors in evaluating the classifier performance.

\subsection{Experimental Results}
\begin{figure}[t!]
	\centering
	\includegraphics[width=1\columnwidth]{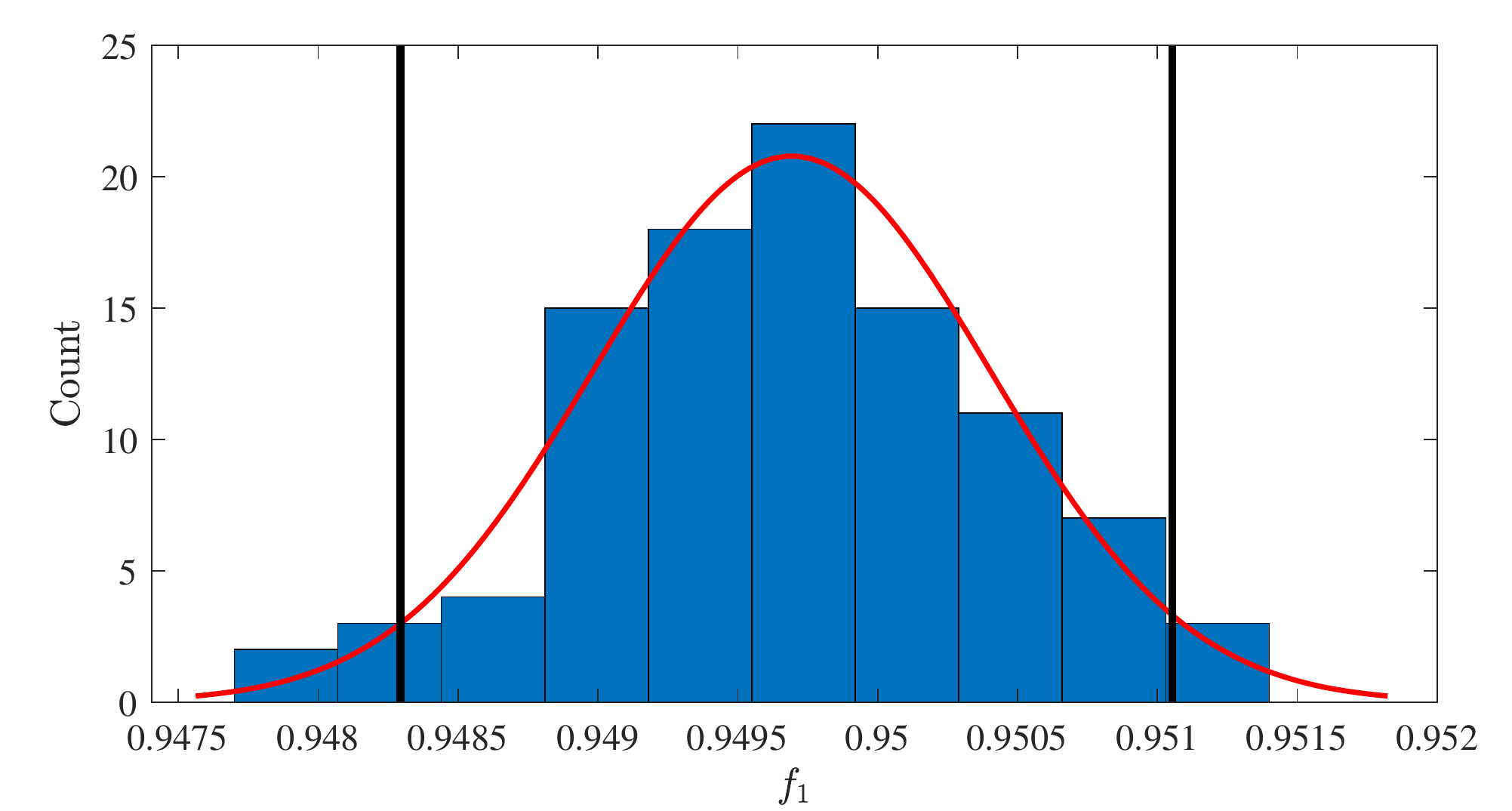}
	\caption{The histogram illustrates the F1-score distribution obtained from DT with the 100 different testing sets. The area in between the black lines indicate the 95\% confidence interval.}
	\label{fig:ciIllustration}
\end{figure}

We fed the two consolidated datasets, i.e., direct and crosswise datasets, to the four different classifiers (i.e., DT, LDA, NB, and kNN) for training.
We repeated the experiment 100 times with a different set of testing data. 
Specifically, we randomly sampled 20\% of data from the dataset for testing purposes at every iteration.
For each evaluation metric, we show the mean result and its corresponding 95\% confidence interval (CI).
An illustration of the F1-score distribution obtained from DT with the 100 testing sets, is shown in Fig.~\ref{fig:ciIllustration}. 
The overall mean results and 95\% CI for both direct and crosswise dataset are shown in Table~\ref{table:performanceS} and Table~\ref{table:performanceO}, respectively.
From both tables, we can see that all the classifiers achieve satisfactory performance with high precision and recall. 
In other words, the classifier did not penalize the recall in order to achieve high precision.
Hence, the F1-scores for both datasets are high.

\begin{figure*}[t!]
	\centering
	\includegraphics[width=0.9\textwidth]{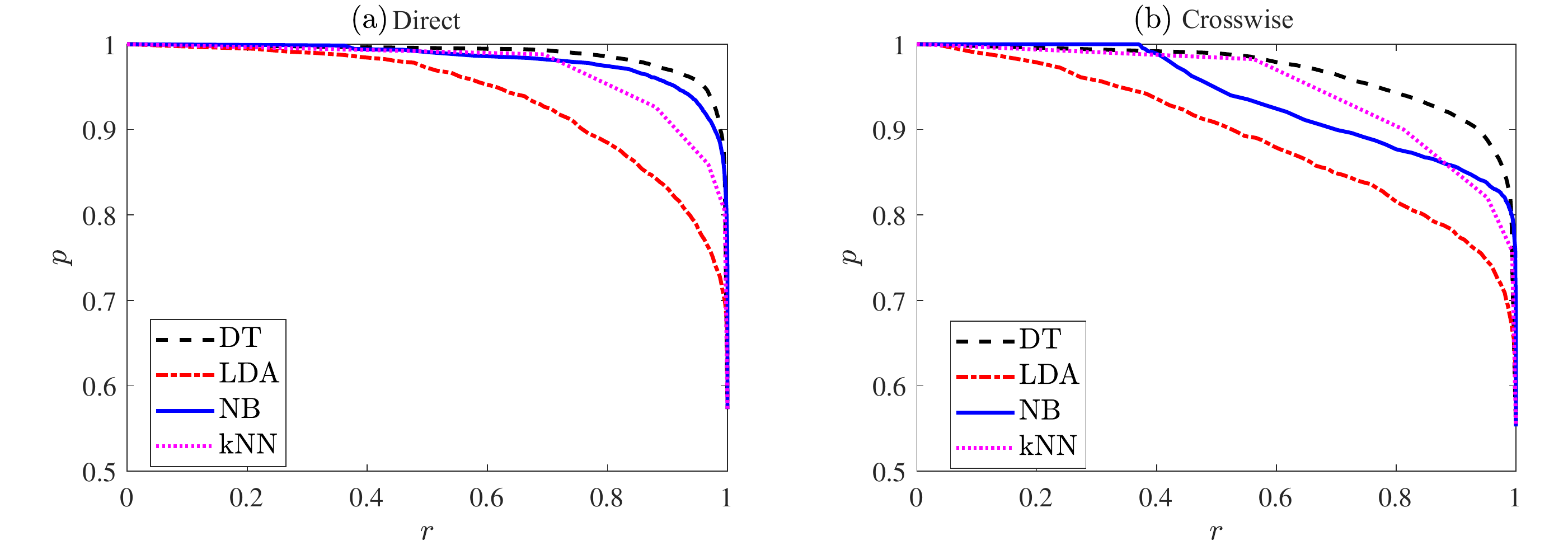}
	\caption{Precision-recall curves for the two types of dataset: (a) direct, (b) crosswise.}
	\label{fig:precisionRecallCurve}
\end{figure*}

\begin{figure*}[t!] 
	\centering
	\includegraphics[width=0.95\textwidth]{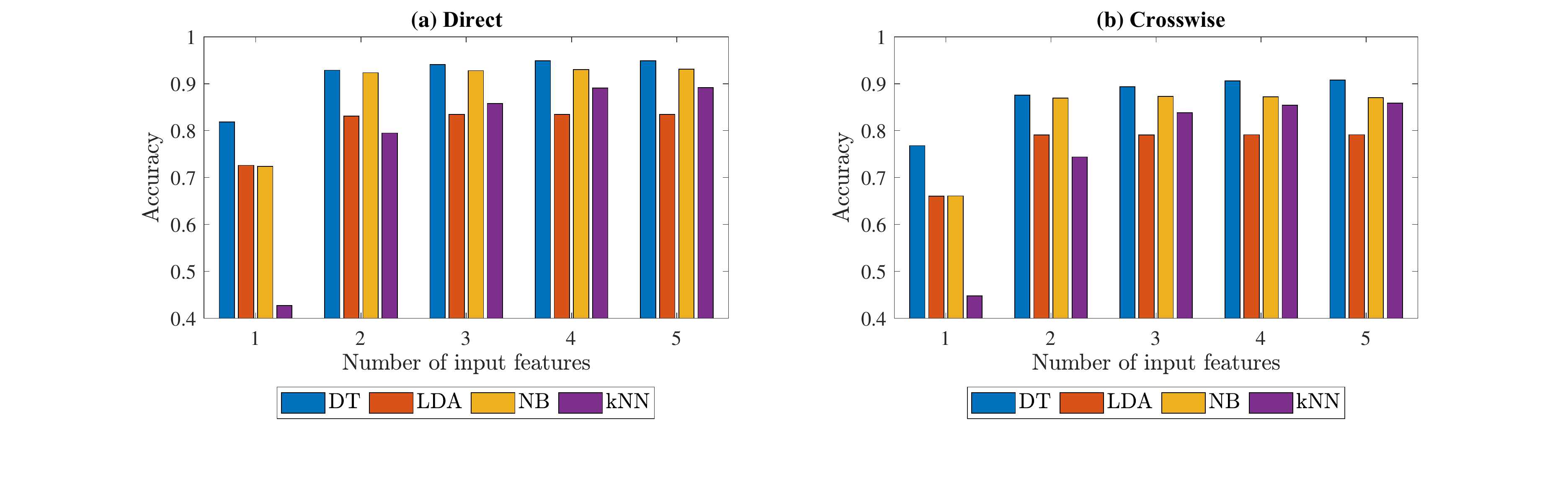}
	\caption{The effect of number of input features on the accuracy.}
	\label{fig:impFeatBar}
\end{figure*}

We also observed that the direct dataset gave a better performance than the crosswise dataset.
This can be explained by the possible signal attenuation when the two hands are blocked by the human body.
Among all the classifiers, DT achieves the best performance with the highest precision, recall, F1-score, and accuracy.
In Fig.~\ref{fig:precisionRecallCurve}, it shows the precision-recall curve for (a) direct and (b) crosswise.
The precision-recall curve provides further insight into the trade-off between precision and recall.
Both plots indicate that DT achieves superior performance with high precision and recall, whereas other methods tend to trade-off the recall in order to achieve high precision.

\begin{figure*}
	\centering
	\includegraphics[width=0.95\textwidth]{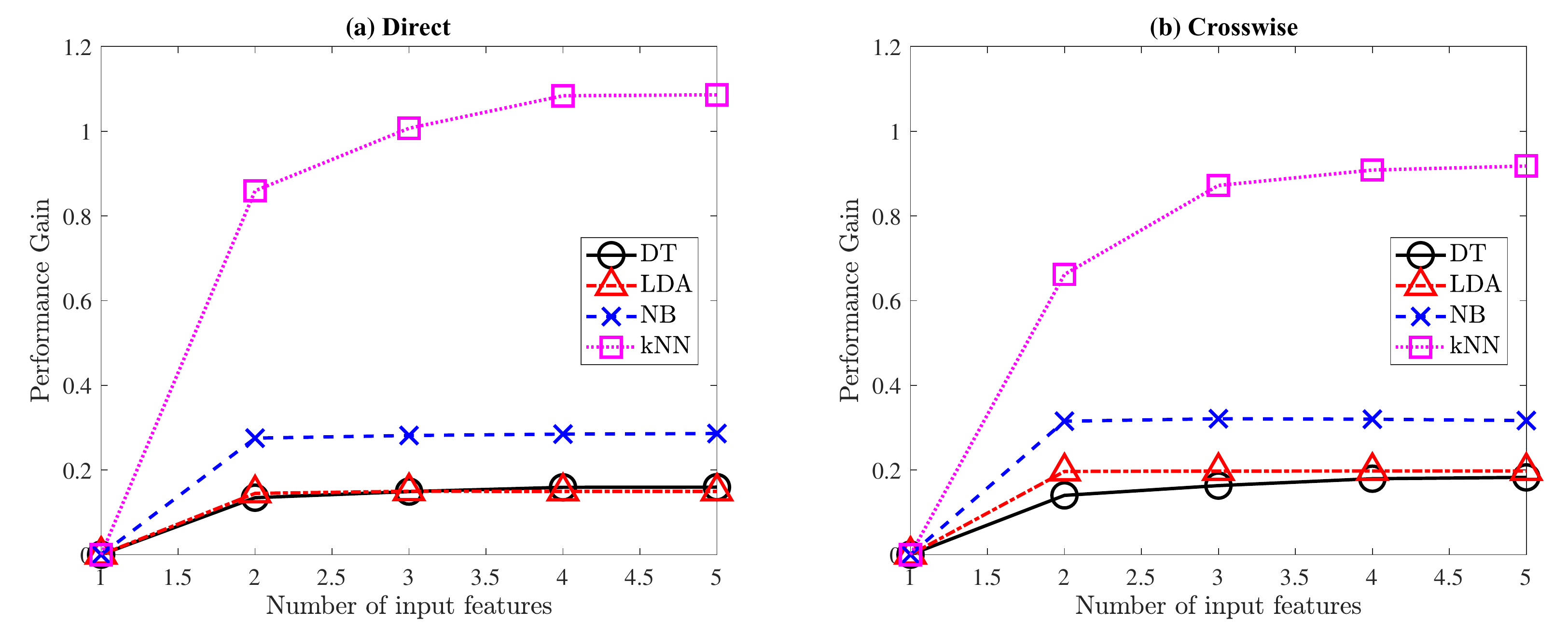}
	\caption{The histogram illustrates the F1-score distribution obtained from DT with the 100 different testing sets. The area in between the black lines indicate the 95\% confidence interval.}
	\label{fig:performanceGain}
\end{figure*}
\begin{figure*}[t!]
	\centering
	\includegraphics[width=0.95\textwidth]{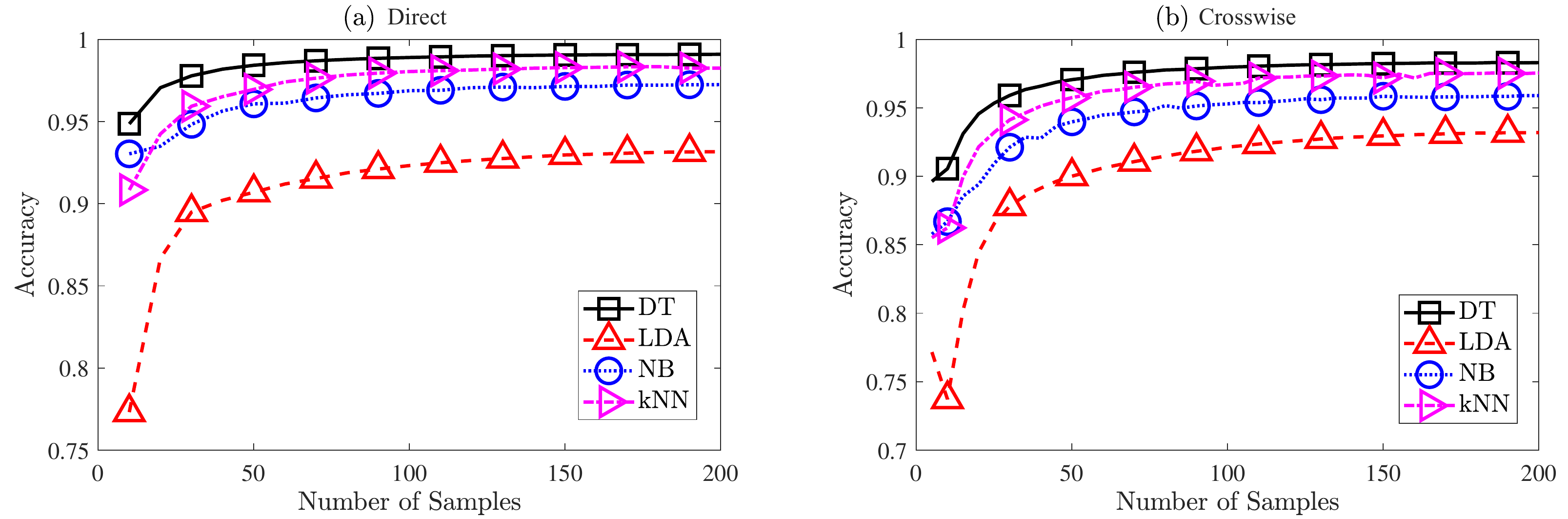}
	\caption{The effect of number of input samples on the accuracy.}
	\label{fig:resSamples}
\end{figure*}

\subsection{Implication of Input Features}
Previously, we used all the five input features (i.e., number of samples observed by the smartwatch, mean RSS, maximum RSS, minimum RSS, and RSS range) to train the model.
All the four trained classifiers were able to produce satisfactory classification performance, i.e., at least 85\% accuracy.
Hence, we would like to investigate the implication of input features on the classification performance.
We repeated the experiment by using only one feature (i.e., mean RSS), and then two features (i.e., mean RSS and the number of samples), and so on.
The classification accuracy achieved by all the four classifiers is shown in Fig.~\ref{fig:impFeatBar}.
From both bar charts, we can see that kNN suffers severe performance degradation when only one input feature available. 
Overall, the performance increases when the number of features increases.

The performance gain of each classifier when the number of features increases, is shown in Fig.~\ref{fig:performanceGain}.
We can see that kNN benefited a lot when there are more input features.
On the other hand, both LDA and NB did not show improvement after two features. 
Their performance saturated when the number of features is more than two.
It can be noted that the performance of DT also increases when the number of features increases, even though the performance gain is quite minimal.
Overall, we can see that some features are indeed useful in training a good model, while some features might be redundant and can be excluded from training.
For example, the maximum RSS and minimum RSS might not provide good information to the model training, whereas the RSS range provides more useful information.
The RSS range provides an indication of how big the RSS fluctuated during a particular observation period, and this piece of information is indeed helpful to model learning.

\subsection{Implication of Number of Samples}
As discussed, the number of samples observed by the smartwatch within a certain continuous time period is a good indication of how long the user has been interacting with each other. 
Furthermore, we can make a better inference when the number of samples observed by the smartwatch increases.
The effect of the number of samples on the classification accuracy is illustrated in Fig.~\ref{fig:resSamples}.
It is clear that the accuracy increases when the number of samples increases and then slowly saturates after it obtains a sufficient number of samples.
In other words, the increase in the number of samples has less effect on accuracy when the system has obtained a sufficient number of samples to make an inference.
From the results, we can see that the accuracy starts to saturate when the number of samples reaches 100, for the (a) direct and (b) crosswise cases.
Hence, we can conclude that most classifiers can produce proper classification output when there are at least 100 samples. 
If the smartwatch is configured to advertise the packet every 100~ms, we should expect approximately 10~samples per second, which means that approximately 10~s are required for each classifier to reach a stable performance.
In practice, this is a reasonable duration considering the interaction duration between users. 
Furthermore, if the interaction duration is less than 10~s, the risk of getting infected is low even if the user is very close to the infected individual.

\subsection{Implication of Physical Distancing Requirements}
\begin{figure*}[t!]
	\centering
	\includegraphics[width=0.95\textwidth]{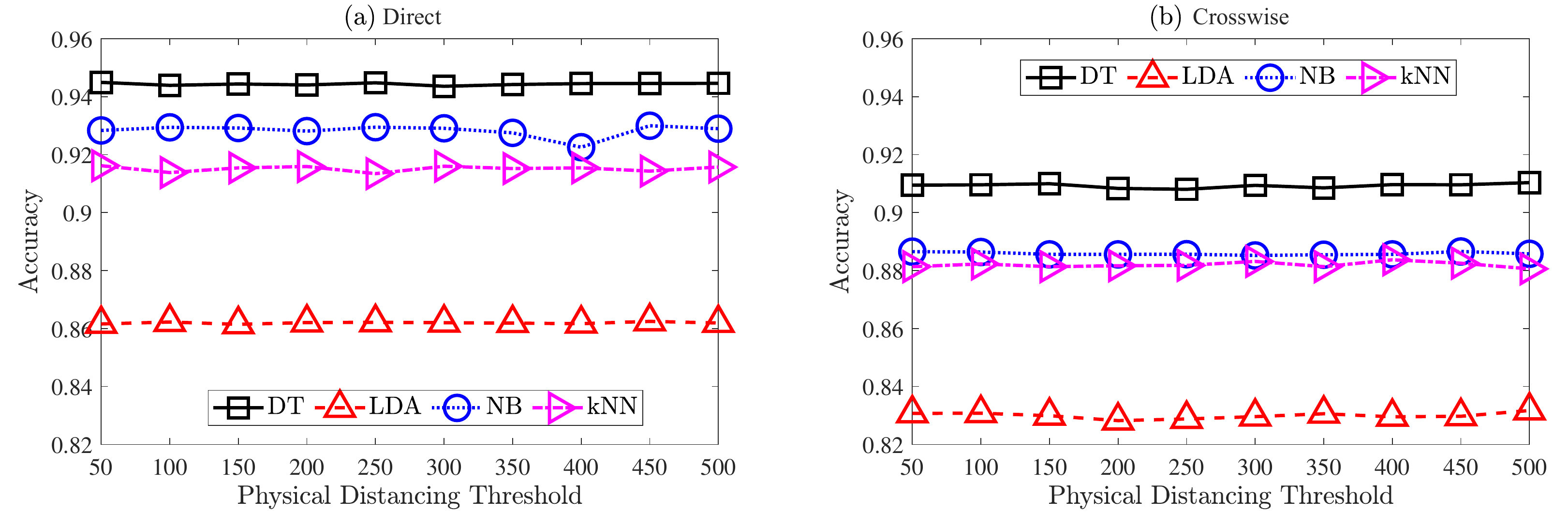}
	\caption{The effect of physical distancing threshold on the accuracy.}
	\label{fig:varyingDistThres2}
\end{figure*}

The World Health Organization recommends a distance of at least one meter. However, different countries implement different physical distancing requirements, from 2 m to 1 m, depending on factors including location, activity, and age of the individuals.
Considered the variations in physical distancing requirements, we conducted an experiment to verify our classification approach with different physical distancing thresholds. The classification accuracy with different physical distancing threshold is shown in Fig.~\ref{fig:varyingDistThres2}.
The results prove the robustness of our classification approach, in which each classifier achieves almost similar accuracy despite the differences in the physical distancing threshold.
This means that our proposed approach is practical and can be applied in any setting directly by simply updating the physical distancing threshold in correspondence to the set of required preventive measures.

\section{Conclusion}
\label{sec:conclusions}
Contact tracing is deemed to be an essential measure in the post-pandemic to prevent the second outbreak while slowly reopening the workplace.
Even though smartphone-based contact tracing is cost-effective considering the ubiquity of smartphones, it is not convenient to have the employee carry with them the smartphone all the time during working.
On the other hand, a smart wearable approach provides a more practical solution to contact tracing in the workplace.
In this paper, we verify the practicality of our proposed P$^3$CT with real-world BLE data collected from the smartwatch.
For future work, we can integrate the embedded sensors within the watch to monitor employee's activity and thus to better predict their interaction behaviors.
The additional knowledge of interaction behaviors, besides the interaction proximity and duration, provide further information to estimate the risk of being infected.



\section*{Acknowledgment}
This project is funded by NSERC Alliance COVID-19 grant \# 552130-20.



%
%
%
 
\bibliographystyle{IEEEtran}
\bibliography{references}

\begin{thebibliography}{10}
\providecommand{\url}[1]{#1}
\csname url@samestyle\endcsname
\providecommand{\newblock}{\relax}
\providecommand{\bibinfo}[2]{#2}
\providecommand{\BIBentrySTDinterwordspacing}{\spaceskip=0pt\relax}
\providecommand{\BIBentryALTinterwordstretchfactor}{4}
\providecommand{\BIBentryALTinterwordspacing}{\spaceskip=\fontdimen2\font plus
\BIBentryALTinterwordstretchfactor\fontdimen3\font minus
  \fontdimen4\font\relax}
\providecommand{\BIBforeignlanguage}[2]{{%
\expandafter\ifx\csname l@#1\endcsname\relax
\typeout{** WARNING: IEEEtran.bst: No hyphenation pattern has been}%
\typeout{** loaded for the language `#1'. Using the pattern for}%
\typeout{** the default language instead.}%
\else
\language=\csname l@#1\endcsname
\fi
#2}}
\providecommand{\BIBdecl}{\relax}
\BIBdecl

\bibitem{ferretti2020quantifying}
L.~Ferretti, C.~Wymant, M.~Kendall, L.~Zhao, A.~Nurtay, L.~Abeler-D{\"o}rner,
  M.~Parker, D.~Bonsall, and C.~Fraser, ``Quantifying sars-cov-2 transmission
  suggests epidemic control with digital contact tracing,'' \emph{Science},
  2020.

\bibitem{eames2003contact}
K.~T. Eames and M.~J. Keeling, ``Contact tracing and disease control,''
  \emph{Proceedings of the Royal Society of London. Series B: Biological
  Sciences}, vol. 270, no. 1533, pp. 2565--2571, 2003.

\bibitem{PEPP}
\BIBentryALTinterwordspacing
Pan-european privacy-preserving proximity tracing. 2020. [Online]. Available:
  \url{https://www.pepp-pt.org/}
\BIBentrySTDinterwordspacing

\bibitem{COVIDwatch}
\BIBentryALTinterwordspacing
We put the power to reduce the spread of covid-19 in the palm of your hand.
  [Online]. Available: \url{https://www.covid-watch.org/}
\BIBentrySTDinterwordspacing

\bibitem{PACT}
\BIBentryALTinterwordspacing
Pact: Private automated contact tracing. [Online]. Available:
  \url{https://pact.mit.edu/}
\BIBentrySTDinterwordspacing

\bibitem{9000599}
P.~C. {Ng}, J.~{She}, and R.~{Ran}, ``A reliable smart interaction with
  physical thing attached with ble beacon,'' \emph{IEEE Internet of Things
  Journal}, vol.~7, no.~4, pp. 3650--3662, April 2020.

\bibitem{chen}
Z.~{Chen}, Q.~{Zhu}, and Y.~C. {Soh}, ``Smartphone inertial sensor-based indoor
  localization and tracking with ibeacon corrections,'' \emph{IEEE Transactions
  on Industrial Informatics}, vol.~12, no.~4, pp. 1540--1549, 2016.

\bibitem{werner}
M.~{Werner}, M.~{Kessel}, and C.~{Marouane}, ``Indoor positioning using
  smartphone camera,'' in \emph{2011 International Conference on Indoor
  Positioning and Indoor Navigation}, 2011, pp. 1--6.

\bibitem{liu}
S.~{Liu}, Y.~{Jiang}, and A.~{Striegel}, ``Face-to-face proximity
  estimationusing bluetooth on smartphones,'' \emph{IEEE Transactions on Mobile
  Computing}, vol.~13, no.~4, pp. 811--823, 2014.

\bibitem{skCorona}
\BIBentryALTinterwordspacing
Coronavirus mobile apps are surging in popularity in south korea. [Online].
  Available:
  \url{https://www.cnn.com/2020/02/28/tech/korea-coronavirus-tracking-apps/index.html}
\BIBentrySTDinterwordspacing

\bibitem{nguyen2020epidemic}
K.~A. Nguyen, Z.~Luo, and C.~Watkins, ``Epidemic contact tracing with
  smartphone sensors,'' \emph{arXiv preprint arXiv:2006.00046}, 2020.

\bibitem{tripathy2020easyband}
A.~K. Tripathy, A.~G. Mohapatra, S.~P. Mohanty, E.~Kougianos, A.~M. Joshi, and
  G.~Das, ``Easyband: A wearable for safety-aware mobility during pandemic
  outbreak,'' \emph{IEEE Consumer Electronics Magazine}, 2020.

\bibitem{chinaCorona}
\BIBentryALTinterwordspacing
China launches coronavirus 'close contact detector' app. [Online]. Available:
  \url{https://www.bbc.com/news/technology-51439401}
\BIBentrySTDinterwordspacing

\bibitem{sgCorona}
\BIBentryALTinterwordspacing
Tracetogether, safer together. [Online]. Available:
  \url{https://www.tracetogether.gov.sg/}
\BIBentrySTDinterwordspacing

\bibitem{shukla2020privacy}
M.~Shukla, S.~Lodha, G.~Shroff, R.~Raskar \emph{et~al.}, ``Privacy guidelines
  for contact tracing applications,'' \emph{arXiv preprint arXiv:2004.13328},
  2020.

\bibitem{bell2020tracesecure}
J.~Bell, D.~Butler, C.~Hicks, and J.~Crowcroft, ``Tracesecure: Towards privacy
  preserving contact tracing,'' \emph{arXiv preprint arXiv:2004.04059}, 2020.

\bibitem{gomez2012overview}
C.~Gomez, J.~Oller, and J.~Paradells, ``Overview and evaluation of bluetooth
  low energy: An emerging low-power wireless technology,'' \emph{Sensors},
  vol.~12, no.~9, pp. 11\,734--11\,753, 2012.

\bibitem{7000963}
K.~{Chang}, ``Bluetooth: a viable solution for iot? [industry perspectives],''
  \emph{IEEE Wireless Communications}, vol.~21, no.~6, pp. 6--7, December 2014.

\bibitem{7366936}
M.~Radhakrishnan, A.~Misra, R.~K. Balan, and Y.~Lee, ``Smartphones and ble
  services: Empirical insights,'' in \emph{2015 IEEE 12th International
  Conference on Mobile Ad Hoc and Sensor Systems}, Oct 2015, pp. 226--234.

\bibitem{8011489}
S.~R. {Hussain}, S.~{Mehnaz}, S.~{Nirjon}, and E.~{Bertino}, ``Secure seamless
  bluetooth low energy connection migration for unmodified iot devices,''
  \emph{IEEE Transactions on Mobile Computing}, vol.~17, no.~4, pp. 927--944,
  April 2018.

\bibitem{8242361}
K.~E. Jeon, J.~She, P.~Soonsawad, and P.~C. Ng, ``Ble beacons for internet of
  things applications: Survey, challenges, and opportunities,'' \emph{IEEE
  Internet of Things Journal}, vol.~5, no.~2, pp. 811--828, April 2018.

\bibitem{8395148}
R.~{Sari} and H.~{Zayyani}, ``Rss localization using unknown statistical path
  loss exponent model,'' \emph{IEEE Communications Letters}, vol.~22, no.~9,
  pp. 1830--1833, Sep. 2018.

\bibitem{8423010}
C.~H. {Lam}, P.~C. {Ng}, and J.~{She}, ``Improved distance estimation with ble
  beacon using kalman filter and svm,'' in \emph{2018 IEEE International
  Conference on Communications (ICC)}, May 2018, pp. 1--6.

\bibitem{7174982}
M.~Leigsnering, F.~Ahmad, M.~G. Amin, and A.~M. Zoubir, ``Compressive
  sensing-based multipath exploitation for stationary and moving indoor target
  localization,'' \emph{IEEE Journal of Selected Topics in Signal Processing},
  vol.~9, no.~8, pp. 1469--1483, Dec 2015.

\bibitem{6856188}
M.~Ayadi and A.~B. Zineb, ``Body shadowing and furniture effects for accuracy
  improvement of indoor wave propagation models,'' \emph{IEEE Transactions on
  Wireless Communications}, vol.~13, no.~11, pp. 5999--6006, Nov 2014.

\bibitem{8588347}
Q.~{Tian}, K.~I. {Wang}, and Z.~{Salcic}, ``Human body shadowing effect on
  uwb-based ranging system for pedestrian tracking,'' \emph{IEEE Transactions
  on Instrumentation and Measurement}, vol.~68, no.~10, pp. 4028--4037, Oct
  2019.

\bibitem{9001059}
P.~{Spachos} and K.~N. {Plataniotis}, ``Ble beacons for indoor positioning at
  an interactive iot-based smart museum,'' \emph{IEEE Systems Journal}, pp.
  1--11, 2020.

\bibitem{8019467}
P.~C. {Ng}, J.~{She}, and S.~{Park}, ``Notify-and-interact: A beacon-smartphone
  interaction for user engagement in galleries,'' in \emph{2017 IEEE
  International Conference on Multimedia and Expo (ICME)}, July 2017, pp.
  1069--1074.

\bibitem{8059756}
P.~C. {Ng}, J.~{She}, and S.~Park, ``High resolution beacon-based proximity
  detection for dense deployment,'' \emph{IEEE Transactions on Mobile
  Computing}, vol.~17, no.~6, pp. 1369--1382, June 2018.

\bibitem{8956048}
A.~{Mackey}, P.~{Spachos}, L.~{Song}, and K.~N. {Plataniotis}, ``Improving ble
  beacon proximity estimation accuracy through bayesian filtering,'' \emph{IEEE
  Internet of Things Journal}, vol.~7, no.~4, pp. 3160--3169, April 2020.

\bibitem{8705339}
P.~C. {Ng}, J.~{She}, and R.~{Ran}, ``A compressive sensing approach to detect
  the proximity between smartphones and ble beacons,'' \emph{IEEE Internet of
  Things Journal}, vol.~6, no.~4, pp. 7162--7174, Aug 2019.

\bibitem{git}
\BIBentryALTinterwordspacing
Rss smartwatch dataset. [Online]. Available:
  \url{https://github.com/pc-ng/rss_smartwatch}
\BIBentrySTDinterwordspacing

\end{thebibliography}

\end{document}